\documentclass[prb,aps,showpacs,twocolumn,superscriptaddress]{revtex4-2}
\usepackage[utf8]{inputenc}
\usepackage[T1]{fontenc}                
\usepackage{amsmath, amsfonts, stmaryrd,dsfont,graphicx,physics,hyperref,bm}
\usepackage[caption=false]{subfig}
\usepackage[capitalise]{cleveref}
\def\mb{\mathbf}
\def\a{\alpha}

\def\k{\kappa}
\def\m{\mu}

\def\s{\sigma} 
\def\6{\partial} 
\def\d{\delta}

\def\t{\tau}
\def\p{\pi}
\def\q{\theta}
\def\z{\zeta}
\def\x{\chi}

\newcommand{\ea}[1]{\begin{align}#1\end{align}}
\newcommand{\eq}[1]{\begin{equation}\begin{split}#1\end{split}\end{equation}}

\def\f{\varphi} 
\def\l{\lambda}

\def\uar{\uparrow}
\def\dar{\downarrow}

\begin{document}
	\title{Non-local magnon transport in a magnetic domain wall wave guide}
	\author{Dion M.F. Hartmann}\email[E-mail adress: ]{d.m.f.hartmann@uu.nl}\affiliation{Institute for Theoretical Physics, Utrecht University, Leuvenlaan 4, NL-3584 CE Utrecht, The Netherlands}
	\author {Andreas Rückriegel}\affiliation{Institut für Theoretische Physik, Universität Frankfurt,
		Max-von-Laue Strasse 1, 60438 Frankfurt, Germany}
	\author {Rembert A. Duine}\affiliation{Institute for Theoretical Physics, Utrecht University, Leuvenlaan 4, NL-3584 CE Utrecht, The Netherlands}\affiliation{Department of Applied Physics, Eindhoven University of Technology, P.O. Box 513, 5600 MB Eindhoven, The Netherlands}
	\begin{abstract}
		Magnetic domain walls function as a wave guide for low energy magnons. In this paper we develop the theory for the non-local transport of these bound magnons through a ferromagnetic insulator that are injected and detected electrically in adjacent normal metal leads by spin-flip scatting processes and the (inverse) spin-Hall effect. Our set-up requires a twofold degeneracy of the magnetic ground state, which we realize by an easy axis and hard axis anisotropy, in the ferromagnetic insulator. This is readily provided by a broad range of materials. The domain wall is a a topologically protected feature of the system and we obtain the non-local spin transport through it. Thereby we provide a framework for reconfigureable magnonic devices.
	\end{abstract}
	\maketitle
	\section{Introduction}
	Magnonic devices, where information is processed and transmitted by the manipulation and control of spin waves, have multiple benefits over electronic devices, ranging from lower power usage to higher processing speeds \cite{khi10,kru10,ser10,len11}.
	To realize the control of spin waves, the magnetic domain wall (DW) has been proposed as a wave guide, because it hosts a soft spin wave mode that propagates along the DW \cite{her04,lan2015spin,San15,Wan15a,Wag16}.
	The added benefit of using DWs is that such a magnonic device is easily reconfigured for instance by applying an external magnetic field or by bulk spin waves \cite{Yan11,Wan15}.
	Furthermore, the DW is a solitary wave which enjoys topological protection from the boundary conditions \cite{raj82}, i.e. between two different magnetic domains there is always a DW along the entire interface between these domains. The DW can be moved, bent and contain Bloch points, but still house the bound spin wave modes after such reconfigurations \cite{San15,Wag16}.
	
	Unlike bulk magnons that usually have an energy gap due to anisotropies and external magnetic fields, the bound magnons on the DW are always gapless, irrespective of microscopic details. As such, they are easily excited at low energies. The gapless nature of the bound magnons as well as their robustness to DW deformations also has an intuitive interpretation as a topological property of the magnetic texture: The DW is a boundary between two topologically distinct phases in which the bulk magnons have opposite polarizations, thus there must be a gapless mode at the interface of the two domains, analogous to the edge states of topological insulators \cite{Kim17}. Physically, this mode corresponds to oscillations of the DW position. Recent simulations have shown that the bound DW modes are also robust against impurities and can be excited well below the bulk energy gap \cite{chen2021narrow}.
	
	This work presents a comprehensive theory of the non-local transport of spin angular momentum between two normal metal (NM) leads that is facilitated by the spin waves bound to the DW. From the well-established notion of the bound magnon \cite{winter1961bloch,San15,lan2015spin}, we go beyond the single mode studies to a model that captures the entire spectrum of the spin waves that are localized at the DW, excluding the bulk magnons.
	We explicitly include both the coupling to the lead electrons as well as the Gilbert damping of the magnetization in the ferromagnetic insulator (FMI), both of which are crucial for the description of realistic experiments. 
	This allows for an experimental verification in a setup similar to recent setups that measure the non-local magnon transport through homogeneously magnetized textures \cite{Cor15,Cor16}.
	Our model is further enriched by an additional anisotropy perpendicular to the dominant anisotropy.
	We analytically obtain a Landauer-Büttiker formula for the transport and numerically demonstrate promising properties of the DW channel. Namely, the DW does not only serve as a wave guide, but also as a wave focus, yielding long magnon relaxation length scales.
	
	Recent theoretical works also show that DWs support superfluid spin transport, but this requires a global $U(1)$ symmetry with respect to rotations of the spin around an easy axis \cite{Kim17}.
	However, it remains a challenge to confirm the superfluid nature of the transport experimentally \cite{yuan2018experimental}, mainly because the physical systems do not respect the symmetry.
	For our model, only a $\mathds{Z}(2)$ symmetry of the magnetic ground state is required, i.e. having a twofold degenerate ground state composing the two magnetic domains separated by the DW. Such a symmetry is easily realized in physical systems with an easy axis and hard axis anisotropy.
		
	In the following we will first set out to discuss the bulk dynamics of the bound magnon in detail. Then we introduce the setup and model for the non-local magnon transport between two leads that we consider and obtain the main results. After demonstrating the relevant results on the relaxation length, we conclude by proposing two potential applications. The appendix to this paper contains more details on the derivation of our theoretical model.
	
	\section{Bound magnons}
	We describe the magnetic texture in spherical coordinates $\mb{n}=\mb{M}/M_\textrm{s}=\mb{\hat{x}}\sin\q\sin\f +\mb{\hat{y}}\cos\q+\mb{\hat{z}}\sin\q\cos\f$, with $M_\textrm{s}$ the saturation magnetization. We start out by considering the classical Lagrangian for the uniaxial ferromagnet
	\eq{
		\label{eq:Lag}
		L_{FM}[\mb{n}]=&\int dV \left(\dot\f\x-\mathcal{H}[\mb{n}]\right);\\
		\qquad
		\mathcal{H}[\mb{n}]=&\frac{1}{2}A(\nabla\mb{n})^2-\frac{1}{2}K_y\left(n_y^2+\k n_z^2\right),
	}
	where $\x=-\hbar s \cos\q$ is the conjugate momentum field to the azimuthal angle field $\f$, with $s$ the spin density,
	$A$ is the exchange stiffness and $\k=K_z/K_y\ll1$ is the ratio of the two easy-axis anisotropy constants.
	The easy $y$-axis anisotropy constant $K_y$ together with the boundary conditions $\mb{n}|_{y=\pm\infty}=\pm\hat{\mb{y}}$ ensure that the ground state for the magnetic texture  is that of a DW. Indeed, the equations of motion give a static solution $\mb{n}_0(\f_0,\q_0)$ with
	\eq{
		\label{eq:dws}
		\q_0=2\arctan\exp\left(\frac{y-y_\textrm{DW}}{\tilde{\l}}\right),
	}
	where $\tilde{\l}\equiv\sqrt{A/K_y(1-\k)}$ is the DW width, which is larger than the exchange length $\l=\sqrt{A/K_y}$, due to the off-axis anisotropy $K_z$. This weaker easy $z$-axis anisotropy enforces a Néel DW, $\f_0=0$, and we assume the system is thin and homogeneous along this axis. 
	Therefore all densities presented here are quantities per area, which are obtained from the volumetric densities by multiplication with the film thickness.
	
\begin{figure}
		\includegraphics[width=\linewidth]{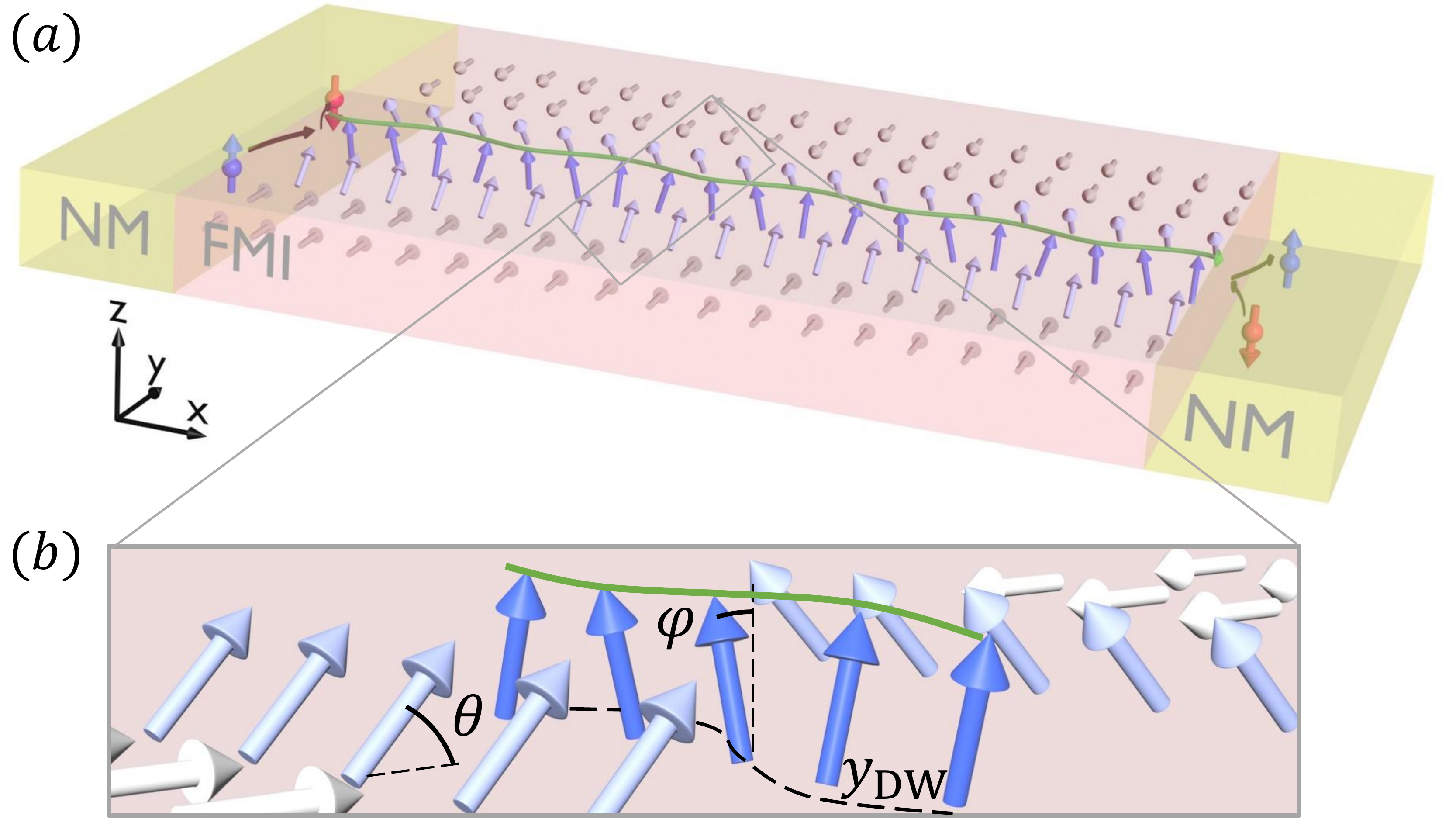}
		\caption{
			\label{fig:sysmdw}
			(Color online)
			(a) A FMI (light red) with a DW texture has an easy $y$- and $z$-axis anisotropies $K_y\ll K_z$. NM leads (light yellow) can inject and absorb magnons due to spin-flip scattering. By applying a bias in the spin population in the left lead, e.g. using the spin Hall effect by applying an electric current in the $y$ direction, there is a net transport of spin angular momentum between the leads. This transport is facilitated by spin waves bound to the DW (green).
			(b) Magnons that are bound to the DW are modeled by oscillations in the azimuthal angle $\f(x,t)$ and the DW position $y_\textrm{DW}(x,t)$ (dashed). Notice that the spin wave displacement (green) along the $y$-axis is proportional to $-y_\textrm{DW}$.
		}
	\end{figure}

	The DW solution is parametrized by its position $y_\textrm{DW}$ and azimuthal angle $\f$. 
	However, its energy is independent of the DW position. 
	Therefore the DW position is a zero mode of the system, and there is a soft mode associated with its long-wavelength fluctuations \cite{raj82}. To capture the dynamics of this mode, we promote the domain wall parameters $y_{\textrm{DW}}$ and $\varphi$ to slowly varying functions of the spatial coordinate $x$ and time $t$. These collective coordinates $y_{\textrm{DW}}(x,t)$ and $\varphi(x,t)$ that describe the low-energy DW dynamics are illustrated in \cref{fig:sysmdw}b.
	Fluctuations in $y_\textrm{DW}$ and $\f$ perturb the magnetization as
	$\d\mb{n}\approx\sin\q_0[\bm{\hat{\f}}\d\f-\bm{\hat{\q}} \d y_\textrm{DW}/\tilde\l]$.
	It is clear, from the prefactor $\sin\q_0$ which vanishes when $|y/y_{\textrm{DW}}|\gg1$, that these are indeed the bound modes.
	
	We expand the FMI Lagrangian in the collective coordinates and evaluate the integral over $y$, yielding 
	\eq{
		L_{FM}=&\int_{FMI} dx\left[-2\hbar s\dot{\f} y_\textrm{DW}\right.
		\\
		&\left.-2K_y\tilde\l\left(\frac{\l^2}{2}(\6_x\f)^2
		+\frac{1-\k}{2}(\6_xy_\textrm{DW})^2
		+\k\f^2
		\right)\right].
	}
	The canonical momentum conjugate to $\f$ is $p=-2\hbar s y_\textrm{DW}$, so we quantize by promoting our collective coordinates to operators with canonical commutation relation $[\f(x),p(x')]=i\hbar\d(x-x')$.
	The bound magnon is then described by the boson field
	\eq{
		\z(x,t)=\begin{pmatrix}
			\eta(x,t) \\
			\eta^\star(x,t)
		\end{pmatrix}; \qquad
		\eta = \sqrt{s \tilde\l}\left(\f-i\frac{y_\textrm{DW}}{\tilde\l}\right).
	}
	The minus sign in the definition of $\eta$ is due to the fact that the spin wave oscillation transverse to the DW, i.e. along the $y$-axis, is proportional to the oscillation of $-y_\textrm{DW}$ as illustrated in \cref{fig:sysmdw}b.
	
	Next, we add Gilbert damping and obtain the equations of motion for our bound magnon field $\zeta$,
	\eq{
		\left\{(i\s_3-\a_G)\hbar\6_t-\frac{K_y}{s}\left[\l^2\6_x^2+\k(1+\s_1)\right]\right\}
	\cdot \zeta = 0,
	}
	where $\a_G$ is the Gilbert damping parameter and $\s_i$ are the Pauli matrices. From the plane wave ansatz $\zeta \sim e^{i(qx-\omega t)}\zeta_\omega$ the dispersion relation follows and
	we obtain four branches determined by solving for $\omega$ from
	\eq{
		\label{eq:disp}
		\l^2q^2(\omega)=-\k+i\a_G\frac{\omega}{\omega_K}\pm\sqrt{\k^2+\frac{\omega^2}{\omega_K^2}},
	}
	with $\omega_K=K_y/\hbar s$ the anisotropic resonance frequency. 
	Here the two branches with the minus sign in front of the square root are exponentially decaying surface states, while the two plus sign branches correspond to bound magnons that propagate forward and backward along the DW. Note also that these propagating modes are indeed gapless, which is due to the fact that the domain wall position is a zero mode. Because of the additional anisotropy $\kappa$, the propagating modes furthermore exhibit a linear dispersion $\omega = v |q|$ at long wavelengths, with a characteristic velocity $v = \sqrt{2 \kappa} / \lambda$.
	
	\section{Non-local magnon transport}
	The set-up we study is illustrated in \cref{fig:sysmdw}a. At the left NM lead there is a net spin bias $\m\equiv\m_\uar-\m_\dar$ which can be generated by the spin Hall effect \cite{Hir99}.
	At the interface with the FMI the exchange interaction between itinerant spins in the NM and spins on the FMI lattice accommodates spin-flip scattering and thereby transfer of spin angular momentum, which excites the bound magnon states of the DW.  
	Gapless magnons then travel to the right interface and can be detected by the reciprocal processes.
	
	The Lagrangian of the FMI-NM interaction is given by
	\eq{
	L_{int}&=-\int_{FM}dV\int_{NM}dV'J_{int}(\mb{r},\mb{r}')s\mb{n}(\mb{r})\cdot\mb{s}(\mb{r}'),
	}
	with $\mb{s}$ the electron spin density of the NM, 
	which we assume to be in a biased steady state, and $J_{int}(\mb{r},\mb{r}')$ the interfacial exchange interaction. Furthermore, we assume the electrons to be non-interacting such that we are able to integrate them out. The Keldysh partition function of the coupled electron-DW magnon system is given by
	\eq{
		Z=&\int\mathcal{D}[\eta,c]e^{\frac{i}{\hbar}\left(S_{FMI}[\eta]+S_{int}[\eta,c]+S_{NM}[c]\right)}\\
		=&\int\mathcal{D}[\eta]e^{\frac{i}{\hbar}S[\eta]},
	}
	where $S$ is the effective action for the magnons on the DW that is obtained by integrating out the leads:
	\eq{
		S=S_{FMI}+\langle S_{int}\rangle_c+\frac{i}{2\hbar}\left(\langle S_{int}^2\rangle_c-\langle S_{int}\rangle_c^2\right)+\mathcal{O}(S_{int}^3),
	}
	with
	\eq{
		\langle \ldots\rangle_c=\int \mathcal{D}[c]\ldots e^{\frac{i}{\hbar}S_{NM}[c]}.
	}
	\begin{figure*}
		\centering
		\subfloat[\label{fig:damp1}]{ \includegraphics[width=0.49 \linewidth]{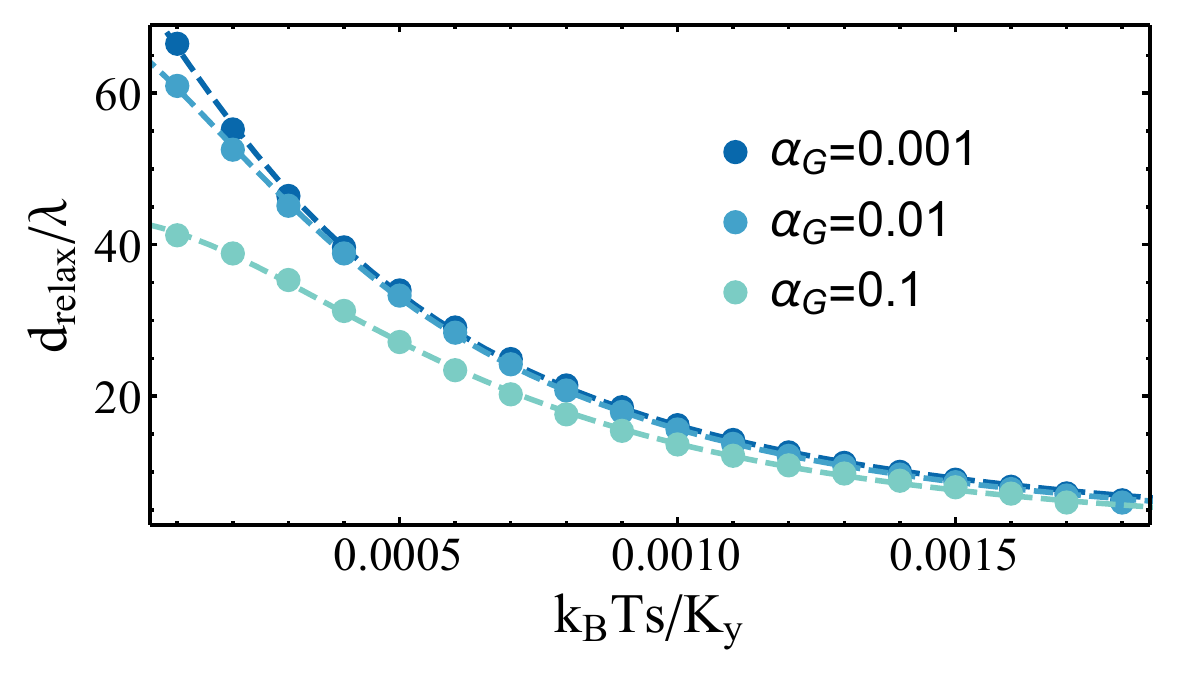}}
		\hfill
		\subfloat[\label{fig:damp2}]{ \includegraphics[width=0.49\linewidth]{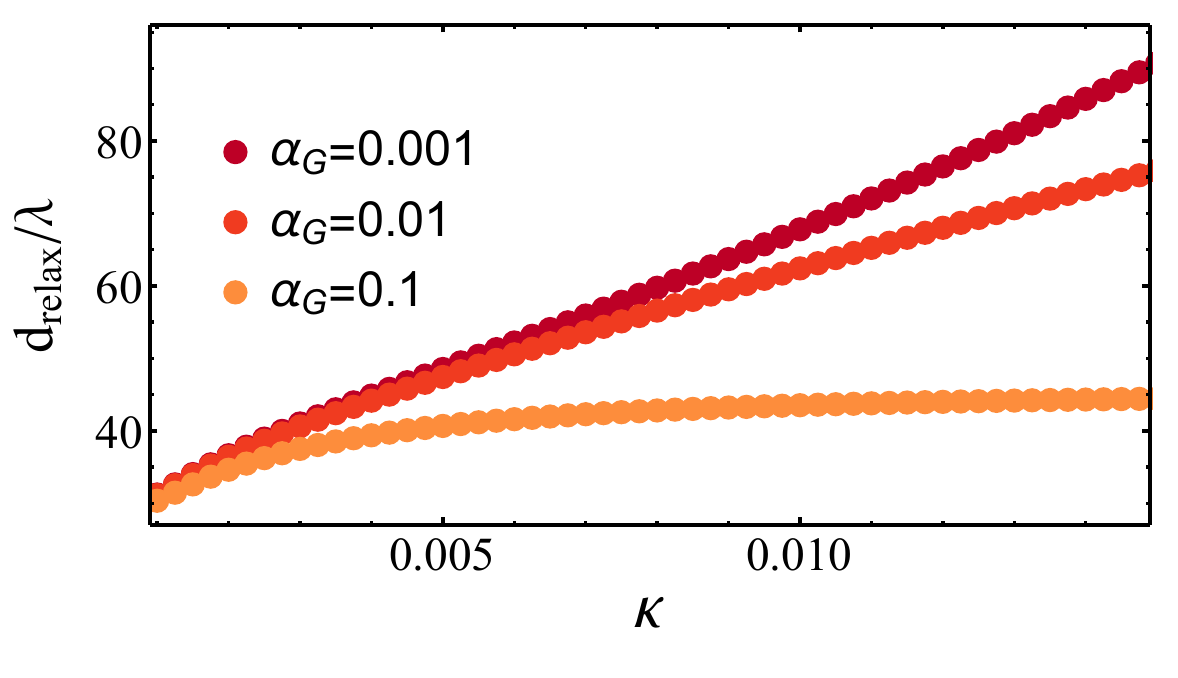}}
		\caption{(Color online) 
			As the distance $d$ between the leads increases, the signal strength decays exponentially, $I_r\sim e^{-d/d_r}$, over a relaxation length scale $d_r$. 
			In (a) the temperature dependence of this length scale is investigated and fitted with $d_r^{-1}=a+b T+c T^2$ for different values of the Gilbert damping parameter $\a_G$ and setting $\k=10^{-2}$ .
			(b) The relaxation length as a function of the ratio between the two anisotropies $\k$ for $k_BTs/K_y=10^{-4}$. 
		}
	\end{figure*}

	Then, after applying a Hubbard-Stratonovich transformation to the Keldysch partition function we obtain a stochastic equation of motion for the Fourier transformed bound magnon field $\z(x,t)=\int \frac{d\omega}{2\p} e^{-i\omega t}\z_\omega(x)$ (see the appendix for more details), 
	\ea{
		\label{eq:EOMzeta}
		\left[\hbar\omega\s_3-H_{DW}-\frac{1}{2\hbar}\s_1\Sigma_\omega\right]\cdot
		\z_\omega=h_\omega,}
	where the Hamiltonian operator is given by $H_{DW}=K_y[\l^2\6^2_x-\k(1+\s_1)]/s$ and the self-energies $\Sigma_\omega = \Sigma_\omega^b +\sum_\s\Sigma^l_{\s,\omega} +\Sigma_\omega^r$ are given by
	\ea{
		i \Sigma^l_{\s,\omega}=&\frac{\a'\hbar}{8}\d(x-x')\d(x)(\hbar\omega+\s\m)
		\\ \nonumber
		&\qquad\times
		\begin{pmatrix}
			\p^2-4 & (\p-2\s)^2 \\
			(\p+2\s)^2 & \p^2-4
		\end{pmatrix};
		\\
		i \Sigma^r_\omega=&\frac{\a'}{4}\d(x-x')\d(x-d)\hbar^2\omega
		\\ \nonumber
		&\qquad\times
		\begin{pmatrix}
			\p^2-4 & \p^2+4 \\
			\p^2+4 & \p^2-4
		\end{pmatrix};
		\\
		i \Sigma^b_\omega=&2\a_G\d(x-x')\hbar^2\omega\s_1,
	}
	with $\s\in\{\uar,\dar\}$ and the effective spin-flip length
	\eq{
		\a'=\frac{\tilde\l g^{\uar\dar}}{2\p s}.
	}  
	Here, $g^{\uar\dar}$ is the spin-mixing conductance \cite{brataas2000finite,tserkovnyak2002enhanced}, which is related to the interfacial exchange coupling $J_{int}$ \cite{Zhe17}. The stochastic field $h_\omega=h^l_\omega+h_\omega^b+h^r_\omega$ satisfies the following fluctuation dissipation rules
	\ea{
		\label{eq:fdt1}
		\langle h^l_\omega h^{l\dagger}_{\omega'}  \rangle=&2\p i\d(\omega+\omega')\sigma_1\sum_\sigma\Sigma^l_{\s,\omega} F^l_{\s,\omega};
		\\
		\label{eq:fdt2}
		\langle h^{r\backslash b}_\omega h^{r\backslash b\dagger}_{\omega'}  \rangle=&2\p i\d(\omega+\omega')\s_1\Sigma^{r\backslash b}_\omega  F^{r\backslash b}_\omega ,
	}
	with $F^l_{\s,\omega}=n_B[(\hbar\omega+\s\m)/k_B T]+1/2$ expressed in terms of the Bose-Einstein distribution function $n_B$, and $F^{r\backslash b}_{\omega}=n_B[\hbar\omega/k_B T]+1/2$, assuming homogeneous temperature. Notice the differences with the regular fluctuation dissipation theorem \cite{Zhe17}: We obtain separate fluctuation dissipation theorems for the two polarizations of the electronic spin as long as there is a finite spin accumulation $\mu$. The reason for this is that the leads are in a biased state and thus not in equilibrium.
	
	The stochastic equation of motion \ref{eq:EOMzeta} can be solved formally in terms of a Green's function $g_\omega(x)$ for the bound magnon,
	\eq{
		\z_\omega(d)=\int dx g_\omega(x)\cdot h_\omega(x).
	}
	While it is possible to obtain this Green's function analytically for our model, it is however rather involved so that we do not show it here. Finally, the spin current into the right lead is obtained in a Landauer-Büttiker form as
	\ea{\label{eq:LBT}
		I_{r}=&
		\int d\omega \sum_\s\mathcal{T}_\s(\omega)\Bigg(F^l_\s(\omega)-F^r(\omega)\Bigg);
		\\
		\mathcal{T}_\s=&
		\frac{1}{16 s\hbar^2}\Re\Bigg\{
		\Tr\Bigg[
		g^\dagger_{-\omega}(0)\cdot
		\begin{pmatrix}
			-1 & 1 \\
			-1 & 1
		\end{pmatrix}
		\\ \nonumber
		& \qquad
		\cdot \Sigma^r_\omega \cdot g_\omega(0) \cdot \s_1\cdot \Sigma^l_{\s,\omega}\Bigg]\Bigg\}.
	}
	See the appendix for more details on the derivation of this result.

	
	To study the long-range spin transport, we investigate the decay of the signal strength as a function of the distance $d$ between the leads for different values of the bulk Gilbert damping $\a_G$. In \cref{fig:damp1} we plot the extracted relaxation length scale as a function of temperature and in \cref{fig:damp2} as a function of the anisotropy ratio $\k$. 
	In Yttrium-Iron-Garnet the DW width $\l$ is estimated to be around $78$nm for typical values of the exchange $A$ and anisotropy energy $K_y$ \cite{Kli14,Sta09}. Hence, we obtain relaxations lengths in the order of micrometers.
	We fit the temperature dependent curves to the function $d_r^{-1}=a+b T+c T^2$, where $b$ is a term we expect from the linear dispersion relation \cite{Cor15,hoffman2013landau} and $a$ and $c$ are corrections due to the spin current contact resistance at the NM|FMI interfaces, and the deviation from the linear dispersion for small $\k$ or large $\omega$. This form is motivated by the general form $d_r\sim v \t_t$ where $v$ is the spin wave speed and $\t_t$ is the total relaxation time obtained from the combination of different relaxation processes.
	
	Remarkably, far from being detrimental to the transport, larger anisotropies $\k$ actually enhance the relaxation length, as shown in \cref{fig:damp2}. This can be explained from the spectrum in \cref{eq:disp} of the bound magnons: For larger $\k$, the propagating modes have a higher velocity $v\propto \sqrt{\kappa}$, and the linear regime where all modes propagate with this velocity becomes more dominant.
	
	An important note regards how the remaining bulk of the magnet responds to the biased leads. For large enough bias, the texture will be distorted \cite{hartmann2020spin}, so we have the restriction $\m\ll\hbar\omega_K/\a'$. Furthermore, the interfacial torques on the texture do not lead to significant a distortion or displacement of the DW, but in fact counteract each other on each side of the DW.

	\section{Application}
	With the theory at hand of non-local transport of ungapped magnons bound to the DW, we propose two types of devices that employ this specific mechanism. The state of the FMI determines the transport and thereby determines the spin angular momentum injected into the right lead. Due to the inverse spin-Hall effect \cite{Dya71a,Hir99}, this will generate an electric current which can be detected as is done in typical non-local magnon transport experiments \cite{Cor15,Cor16b,Hua12}. 
	\begin{figure}[b]
	\includegraphics[width=\linewidth]{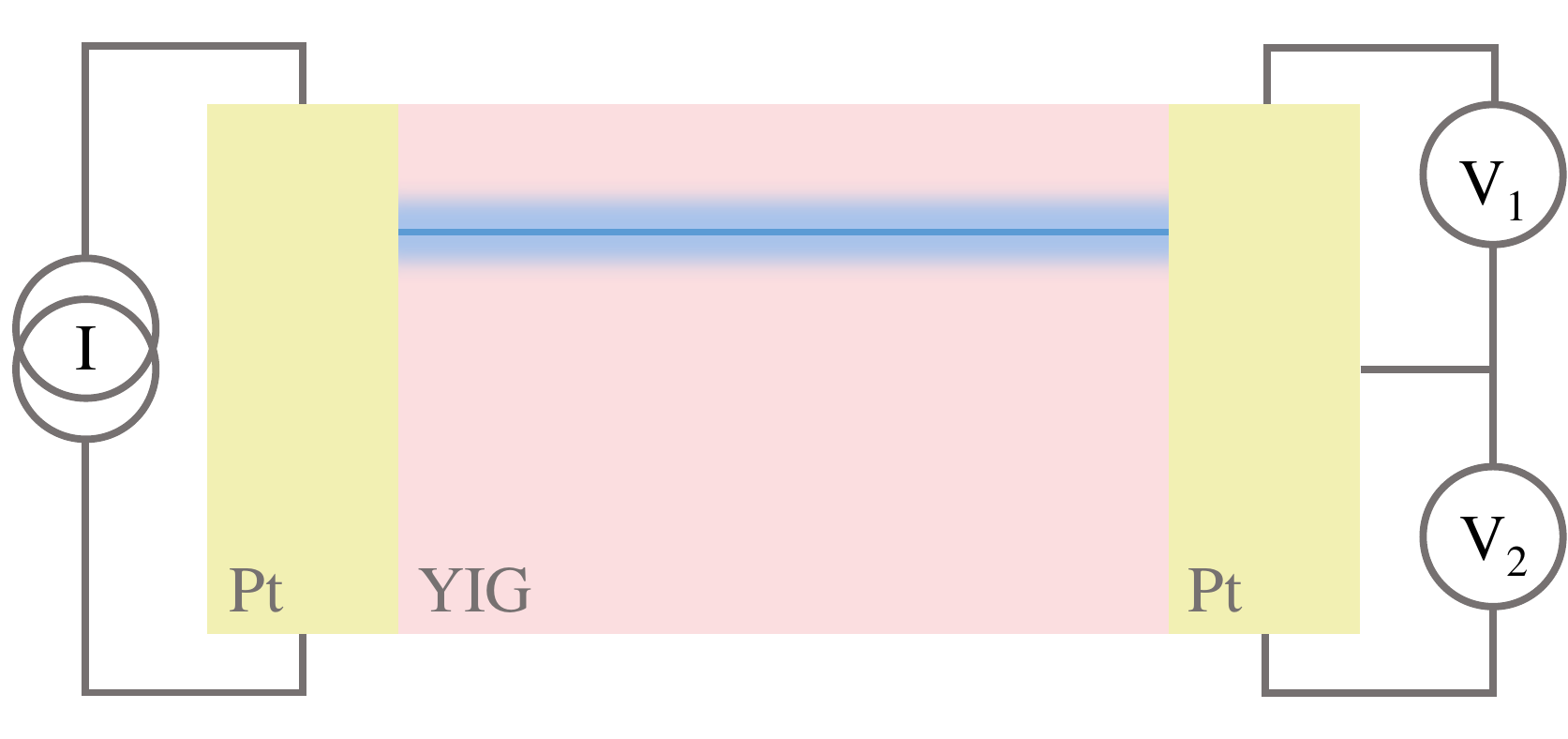}
	\caption{
		\label{fig:switch}
		(Color online)
		The output signals ($V_1$ and $V_2$) are switched on by configuring the DW (blue): When the DW position is in the top half, as illustrated, there will be a finite voltage measured by $V_1$ due to the inverse spin-Hall effect, whereas there will be no signal measured by $V_2$. The proposed materials used are platinum (Pt) for the normal metal leads, and Yttrium-Iron-Garnet (YIG) for the FMI as these are commonly used in non-local magnon transport experiments \cite{Cor16b}.
	}
	\end{figure}
	
	The first type that we propose is a configurable switching device, which is illustrated in \cref{fig:switch}. In the left lead an input spin bias, e.g. by means of an electric current employing the spin-Hall effect, sends a signal to one of the two detectors on the right lead. The position of the DW determines which and can be configured by an external magnetic field aligned with one of the magnetic domains.
	\begin{figure}
		\includegraphics[width=\linewidth]{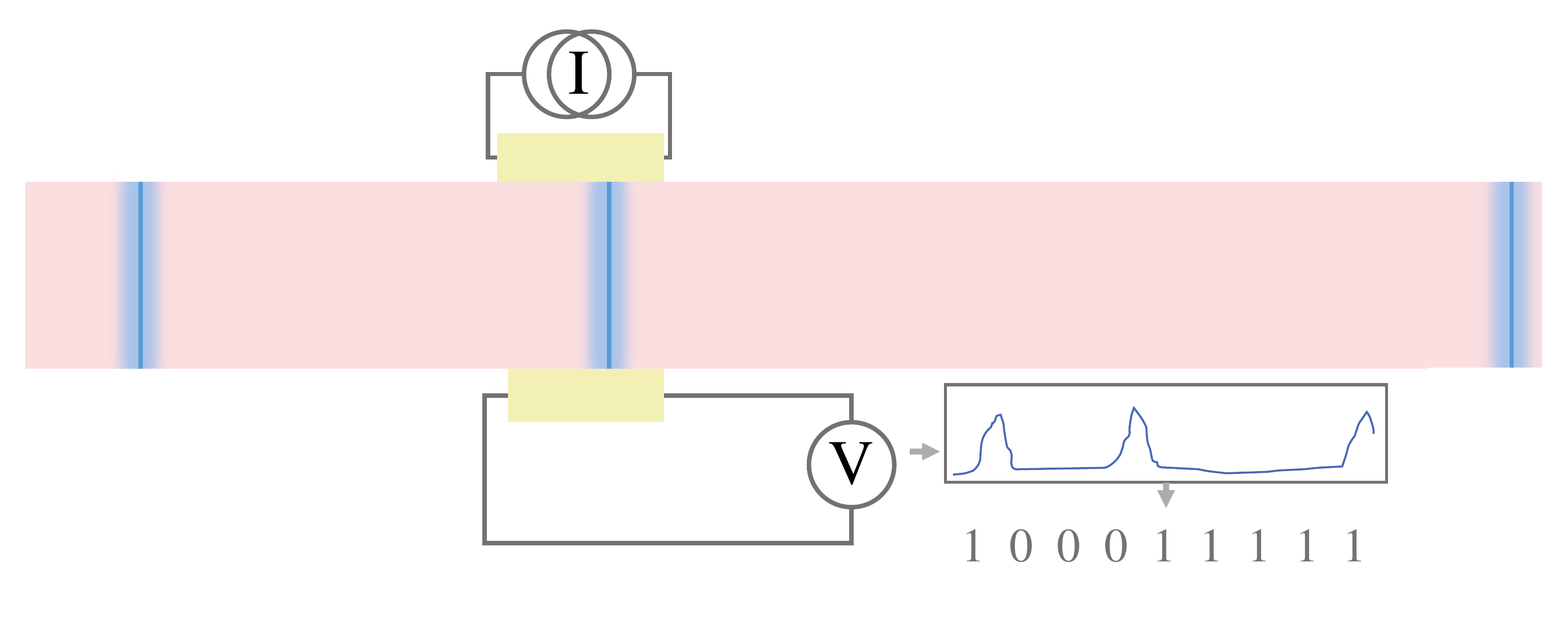}
		\caption{
			\label{fig:dwc}
			(Color online)
			Instead of reading out the orientation of the magnetic domains, this device measures whether a DW (blue) passes and from there infers the data stored on the magnetic track.
		}
	\end{figure}
	
	The second application we propose is a device that counts magnetic domains by counting the DWs. Data stored as magnetic domains, for instance on the racetrack memory \cite{Par08}, is read out by measuring the orientation of the magnetization. But by attaching two leads to such a strip of magnetic domain data as illustrated in \cref{fig:dwc}, one can read out when a DW passes the leads. Then, the counting of 0's and 1's is just a matter of keeping track of when a DW passes. A possible restriction to the DW speed might come from the velocity of the bound spin waves, which has to be significantly larger than the DW velocity.
	
	Magnetic storage devices often use ferromagnetic materials, because the data can be easily read out using the local magnetic field generated by the magnetic domains. Although anti ferromagnets have many benefits over ferromagnetic devices - for instance DWs move much faster - they are so far not employed as materials for storage devices, because it is hard to distinguish and read out the domains. Measuring the anti ferromagnetic DWs instead circumvents this problem and might pave a way for the introduction of anti ferromagnetic materials to the playfield of magnetic storage devices. An extension of our theoretical model to an anti ferromagnetic texture will benefit this path.
	
	\section{Conclusion}
	The ferromagnetic texture with a DW allows two types of spin waves: gapped magnons that propagate through the bulk and ungapped bound magnons that reside on the DW. We have developed a theoretical model for the non-local transport of magnons that is accommodated on DWs. We extracted the transport of spin angular momentum between two normal metal leads by treating the bound spin waves as oscillations of the collective coordinates of the DW solution (see \cref{fig:sysmdw}b), which explicitly exclude the bulk magnons. 
	
	The resulting framework leads to an analytical expression of the transport that takes the familiar Landauer-Büttiker form given in \cref{eq:LBT}, which leads to the intuitive interpretation: The DW connects two baths which have a relative difference in spin population. This bias thus results in a net flow of spin angular momentum. Moreover, there are no significant requirements on our considered system, such as $U(1)$ symmetry, that would make the setup hard to realize experimentally.
	On the contrary, our results show that the breaking of U(1) symmetry is actually beneficial to the transport, because it leads to longer relaxation lengths.
	Our numerical results show furthermore that the obtained transport enjoys benefits from the DW guiding in the form of a large relaxation length. We recognize these features as benefits from the linear and gapless dispersion relation. However, an analytic understanding of the relaxation length is still an open issue.
	
	We expect these DW wave guides to be very versatile and (topologically) robust \cite{San15}. For example, we expect that our theory is also applicable to curved DWs \cite{San15,Wan15a}, as long as the radius of curvature is large compared to the domain wall thickness. Moreover, even in the presence of a pinning potential, caused by impurities in real materials, the bound magnons can still be exited well below the bulk magnon energy gap \cite{chen2021narrow}. Based on our findings, we proposed two types of devices that employ the DW bound magnons to switch a signal by configuring the DW and to read magnetic memory data. Future research should first focus on the experimental verification of our theoretical results before investigating the realizability of these devices. Theoretical progress is yet to be made in extending our model to anti ferromagnets. 
	
	\begin{acknowledgments}
		R.D. is a member of the D-ITP consortium, a program of the Dutch Organization for Scientific Research (NWO) that is funded by the Dutch Ministry of Education, Culture and Science (OCW). R.D. and D.H. have received funding from the European Research Council (ERC) under the
		European Unions Horizon 2020 research and innovation programme (Grant agreement No. 725509).
		This work is part of the research programme of the Foundation for Fundamental Research on Matter (FOM), which is part of the Netherlands Organization for Scientific Research (NWO).	
	\end{acknowledgments}

	\appendix*
	\section{Spin transport formalism}
	In this appendix we give some more details on the derivation of the spin transport from the action of the total system. The full action for our system consists of three parts: the ferromagnetic insulator (FMI), the interaction (int) and the normal metal (NM) leads. The Lagrangian part of the FMI is given in the main text. We assume the action of the electrons in the NM leads to be quadratic in the fermionic operators and diagonal in the spin label. The Lagrangian of the interaction part is given by the exchange interaction between spins from the FMI and the NM,
	\begin{widetext}
	\eq{
		L_{int}&\approx-\tilde\lambda s \p\int_{FMI}dx\d(x)\left(s_x(0)\f-\frac{1}{2} s_z(0)\f^2+2s_y(0)\frac{y_{\textrm{DW}}}{\p\tilde\l}\right).
	}
	Here, we assumed a coarse grained, long wavelength coupling $J_{int}(\mb{r},\mb{r}')=J_{int}\d(\mb{r}-\mb{r}')\d(x)$, expanded in the domain wall collective coordinates $\f$ and $y_{\textrm{DW}}$, and dropped constant terms. We expand the FMI Lagrangian in the collective coordinates
	\eq{
		L_{FMI}=&\int_{FMI} dx\left[-2\hbar s\dot{\f}_0 y_{\textrm{DW}}-2K_y\tilde\l\left(\frac{\l^2}{2}(\6_x\f)^2
		+\frac{1-\k}{2}(\6_xy_{\textrm{DW}})^2
		+\k\f^2
		\right)\right].
	}
	With $p$ the cannonical momentum conjugate to $\f$ we write the Hamiltonian as follows:
	\eq{
		H_{int}+H_{FMI}=\tilde\lambda \int_{FMI}dx \left[s\p \d(x)\left(s_x\f-\frac{1}{2} s_z\f^2+2s_y\frac{y_{\textrm{DW}}}{\p\tilde\l}\right)+2K_y\left(\frac{\l^2}{2}(\6_x\f)^2
		+\frac{1-\k}{2}(\6_xy_{\textrm{DW}})^2
		+\k\f^2
		\right)\right].
	}
	Now we quantize by promoting our coordinates to operators with:
	\ea{
		\left[\f(x),p(x')\right]&=i\hbar\d(x-x'),\\
		\mb{s}&=\frac{1}{2}\left[(c_\uar^\dagger c_\dar+ c_\dar^\dagger c_\uar)\hat{\mb{x}}+i(c_\uar^\dagger c_\dar- c_\dar^\dagger c_\uar)\hat{\mb{y}}+(c_\uar^\dagger c_\uar- c_\dar^\dagger c_\dar)\hat{\mb{z}}\right],
	}
	where $c^\dagger_\sigma$ creates an electron with spin $\sigma$.
	We express the Hamiltonian $H=H_{int}+H_{FMI}$ in terms of the boson field $\eta$ as
	\eq{
		H=&
		\int_{FMI}dx \left[\p \d(x)\left(\sqrt{s\tilde\l}s_x\frac{1}{2}(\eta+\eta^\dagger)-\frac{1}{8} s_z(\eta+\eta^\dagger)^2+\frac{i}{\p}s_y\sqrt{s\tilde\l}(\eta-\eta^\dagger)\right)\right.
		\\
		&\left.+\frac{K_y}{s}\left(\l^2(\6_x\eta)(\6_x\eta^\dagger)
		+\frac{1}{2}\k(\eta+\eta^\dagger)^2\right)\right].
	}
	Now we insert the DW solution into $\mb{S}=-s\hat{\mb{n}}$ and introduce $n_{\uar\backslash\dar}=n_x\pm i n_y$. Then we write 
	\eq{
		S_{int}=-\frac{s\hbar}{2}\sum_{p\in\pm}p\int dt\int_{FMI}dV\int_{NM}dV'J_{int}(\mb{r},\mb{r}')\left[c_\uar^{p\dagger} c^p_\dar n^p_\uar+c_\dar^{p\dagger} c^p_\uar n^p_\dar+\left(c_\uar^{p\dagger} c^p_\uar -c_\dar^{p\dagger} c^p_\dar\right)n^p_z\right].
	}
	Where $p$ indicates which part of the Keldysh contour we evaluate. The electronic Green's functions are defined as
	\eq{
		i G^{pp'}_{\s\s'}=\langle c_\s^{p\dagger} c_{\s'}^{p'} \rangle_c.
	}
	We assume that the leads are in a biased stationary state and neglect the influence of the interfacial coupling $J$ on the electrons, so the Green's functions depend only on the difference in coordinates. Because the lead action is diagonal in spin label, we have that $G^{pp'}_{\s\s'}=0$ when $\s\neq\s'$. So
	\eq{
		\langle S_{int}\rangle_c=i\frac{\p\hbar}{16}\sum_{p,p'\in\pm}pp'\int dtdt'\int dxdx' J_{int}\d(x)\d(x')p'\d_{p,p'}\sum_{\s\in\uar\backslash\dar}\s G^{pp}_{\s} [\zeta^p]^T\begin{pmatrix}
			1 & 1 \\
			1 & 1
		\end{pmatrix}
		\zeta^{p'}
		.
	}
	\eq{
		\langle S_{int}\rangle_c^2=&-\frac{s^2\p^2\tilde\l^2\hbar^2}{4}\sum_{p,p'\in\pm}pp'\int dtdt'\int dx dx' J_{int}^2\d(x)\d(x')
		\\
		&\sum_{\s\in\uar\backslash\dar}\left(G^{pp}_\s G^{p'p'}_\s-G^{pp}_\s G^{p'p'}_{\bar{\s}} \right)\cos\f^p\cos\f^{p'}.
	}
	And using Wick's theorem 
	\eq{
		\langle c^{p_1\dagger}_{\sigma_1}c^{p_2}_{\sigma_2}c^{p_3\dagger}_{\sigma_3}c^{p_4}_{\sigma_4} \rangle=G^{p_4 p_1}_{\sigma_4\sigma_1}G^{p_2 p_3}_{\sigma_2\sigma_3}-G^{p_2 p_1}_{\sigma_2\sigma_1}G^{p_4 p_3}_{\sigma_4\sigma_3}
	}
	we find
	\eq{
		\langle S_{int}^2\rangle_c=&\frac{s^2\hbar^2}{4}\sum_{p,p'\in\pm}pp'\int dtdt'\int dx dx' J_{int}^2\d(x)\d(x')
		\\
		&\times\sum_{\s\in\uar\backslash\dar}\left[
		G^{p'p}_\s G^{pp'}_{\bar{\s}}(\tilde\l\p\f^p+2i\s y_{\textrm{DW}}^p)(\tilde\l\p\f^{p'}-2i\s y_{\textrm{DW}}^{p'})
		\right.
		\\
		&\left.+\left(-G^{pp}_\s G^{p'p'}_\s+G^{pp}_\s G^{p'p'}_{\bar{\s}}+G^{pp'}_\s G^{p'p}_\s\right)\p^2\tilde\l^2\cos\f^p\cos\f^{p'}\right],
	}
	such that
	\eq{
		\langle S_{int}^2\rangle_c-\langle S_{int}\rangle_c^2
		=&\frac{s\tilde\l\hbar^2 }{16}\sum_{p,p'\in\pm}pp'\int dtdt'\int dx dx' J_{int}^2\d(x)\d(x')
		\\
		&\times\sum_{\s\in\uar\backslash\dar}[\zeta^p]^T\left[
		G^{p'p}_\s G^{pp'}_{\bar{\s}}
		\begin{pmatrix}
			\p^2-4 & (\p-2\s)^2 \\
			(\p+2\s)^2 & \p^2-4
		\end{pmatrix}
		\right.
		\\
		&\left.+(G^{p\bar{p}}_\s G^{\bar{p}p}_\s-G^{pp}_\s G^{pp}_\s)\d_{p,p'}\p^2
		\begin{pmatrix}
			1 & 1 \\
			1 & 1
		\end{pmatrix}\right]\zeta^{p'}.
	}
	Therefore the correction to the DW magnon action due to the interaction with the electrons is given by
	\eq{
		\Delta S=& S-S_{FMI}
		=\langle S_{int}\rangle_c+\frac{i}{2\hbar}\left(\langle S_{int}^2\rangle_c-\langle S_{int}\rangle_c^2\right)
		\\
		\approx& 
		\frac{i\hbar}{16}\sum_{p,p'\in\pm}pp'\sum_{\s\in\uar\backslash\dar}\int dtdt'\int dxdx' J_{int}\d(x)\d(x')[\zeta^p]^T
		\left\{\p p'\s G^{pp}_{\s} \begin{pmatrix}
			1 & 1 \\
			1 & 1
		\end{pmatrix}\right.\\
		&\left. +\frac{s\tilde\l}{2}J_{int}\left[G^{p'p}_\s G^{pp'}_{\bar{\s}}
		\begin{pmatrix}
			\p^2-4 & (\p-2\s)^2 \\
			(\p+2\s)^2 & \p^2-4
		\end{pmatrix}
		+(G^{p\bar{p}}_\s G^{\bar{p}p}_\s-G^{pp}_\s G^{pp}_\s)\d_{p,p'}\p^2
		\begin{pmatrix}
			1 & 1 \\
			1 & 1
		\end{pmatrix}
		\right]
		\right\}
		\zeta^{p'}.
	}
	Next, we perform the Keldysh rotation
	\eq{
		\z^p=\frac{1}{\sqrt{2}}(\z^c+p\z^q);\qquad G^{pp'}=\frac{1}{2}(p'G^R+p G^A+G^K),
	}
	where $\zeta^c$ is the classical field, $\zeta^q$ is the quantum field, $G^R$ is the retarded, $G^A$ the advanced and $G^K$ the Keldysh  Green's function.
	Then we have
	\eq{
		\Delta S =-\frac{1}{2}\int dt dt'\int dx dx'\left\{[\z^c]^T\Sigma^A\z^q+[\z^q]^T\Sigma^R\z^c+[\z^q]^T\Sigma^K\z^q\right\},
	}
	with
	\ea{
		\Sigma^{R\backslash A}(x,x',t,t')=&-\frac{i}{16}\hbar J_{int}\d(x-x')\d(x)\sum_{\s\in\uar\backslash\dar}\left\{\d(t-t')\p\s G^K_\s(0,0)
		\begin{pmatrix}
			1 & 1 \\
			1 & 1
		\end{pmatrix}\right.
		\\
		\nonumber
		&\left.
		+\frac{1}{2}\tilde\l s J_{int}\left[\left(G^{R\backslash A}_{\bar{\s}}(0,t-t') G^K_{\s}(0,t'-t)+G^K_{\bar{\s}}(0,t-t') G^{A\backslash R}_{\s}(0,t'-t)\right)
		\begin{pmatrix}
			\p^2-4 & (\p-2\s)^2 \\
			(\p+2\s)^2 & \p^2-4
		\end{pmatrix}
		\right.\right.
		\\
		\nonumber
		&\left.\left.
		-\p^2\d(t-t')\left(G^{R}_\s(0,0) G^K_\s(0,0)+G^K_\s(0,0) G^{A}_\s(0,0)\right)
		\begin{pmatrix}
			1 & 1 \\
			1 & 1
		\end{pmatrix}
		\right]
		\right\},
		\\
		\Sigma^K(x,x',t,t')=&-\frac{i}{32}\tilde\l s\hbar J_{int}^2\d(x-x')\d(x)	
		\begin{pmatrix}
			\p^2-4 & (\p-2\s)^2 \\
			(\p+2\s)^2 & \p^2-4
		\end{pmatrix}
		\\
		\nonumber
		&
		\sum_{\s\in\uar\backslash\dar}
		\left(G^A_{\bar{\s}}(0,t-t') G^R_\s(0,t'-t)+G^R_{\bar{\s}}(0,t-t') G^A_\s(0,t'-t)+G^K_{\bar{\s}}(0,t-t') G^K_\s(0,t'-t)\right).
	}
	Now we Fourier transform
	\eq{
		\Sigma(t,t')=\int\frac{d\omega}{2\p}e^{-i\omega(t-t')}\Sigma(\omega),\qquad \Sigma(\omega)=\int dte^{i\omega t} \Sigma(t,0), \qquad
		G(t)=\int\frac{d\omega}{2\p}e^{-i\omega t}G(\omega),
	}
	to obtain
	\ea{
		\Sigma^{R\backslash A}(x,x';\omega)=&-\frac{i\hbar}{16}J_{int}\d(x-x')\d(x)\sum_{\s\in\uar\backslash\dar}\left\{\p\s \int\frac{d\omega'}{2\p}G^K_\s(0,\omega')
		\begin{pmatrix}
			1 & 1 \\
			1 & 1
		\end{pmatrix}\right.
		\\
		\nonumber
		&\left.
		+\frac{1}{2}\tilde\l s J_{int}\left[\int\frac{d\omega'}{2\p}\left(G^{R\backslash A}_{\bar{\s}}(0,\omega') G^K_{\s}(0,\omega'-\omega)+G^K_{\bar{\s}}(0,\omega') G^{A\backslash R}_{\s}(0,\omega'-\omega)\right)
		\begin{pmatrix}
			\p^2-4 & (\p-2\s)^2 \\
			(\p+2\s)^2 & \p^2-4
		\end{pmatrix}
		\right.\right.
		\\
		\nonumber
		&\left.\left.
		-\p^2\d(t-t')\int\frac{d\omega'}{2\p}\frac{d\omega''}{2\p}\left(G^{R}_\s(0,\omega') G^K_\s(0,\omega'')+G^K_\s(0,\omega') G^{A}_\s(0,\omega'')\right)
		\begin{pmatrix}
			1 & 1 \\
			1 & 1
		\end{pmatrix}
		\right]
		\right\},
		\\
		\Sigma^K(x,x';\omega)=&-\frac{i}{32}\tilde\l s\hbar J_{int}^2\d(x-x')\d(x)
		\begin{pmatrix}
			\p^2-4 & (\p-2\s)^2 \\
			(\p+2\s)^2 & \p^2-4
		\end{pmatrix}
		\\
		\nonumber
		&
		\sum_{\s\in\uar\backslash\dar}\int\frac{d\omega'}{2\p}
		\left(G^A_{\bar{\s}}(0,\omega') G^R_\s(0,\omega'-\omega)+G^R_{\bar{\s}}(0,\omega') G^A_\s(0,\omega'-\omega)+G^K_{\bar{\s}}(0,\omega') G^K_\s(0,\omega'-\omega)\right).
	}
	In the next step, we write
	\eq{
		G^{R\backslash A}_\s(0,\omega')=\lim_{\epsilon\dar 0}\int\frac{d\omega''}{2\p}\frac{A(\omega'')}{\omega'-\omega''\pm i\epsilon},
	}
	where $A(\omega)$ is the electronic spectral function at the interface. It is independent of the spin label $\sigma$ because we are considering normal metal leads. Then, using the following identity
	\eq{
		\d(\omega-\omega')=\lim_{\epsilon\dar 0}\frac{\epsilon}{\p}\frac{1}{(\omega-\omega')^2+\epsilon^2},
	}
	we can write
	\eq{
		G^K_\s(\omega')=\tanh\left(\frac{\hbar\omega'-\m_\s}{2k_B T} \right)\left[G^R_\s(\omega')-G^A_\s(\omega')\right]=-i A(\omega')\tanh\left(\frac{\hbar\omega'-\m_\s}{2k_B T}
		\right).
	}
	Now, assuming $|\m_\sigma|,|\omega|,k_B T\ll\varepsilon_F$, we expand
	\eq{
		\tanh\left(\frac{\hbar\omega'-\m_{\bar{\s}}}{k_B T}\right)-\tanh\left(\frac{\hbar(\omega'-\omega)-\m_{\s}}{k_B T}\right)\approx2\d(\hbar\omega'-\varepsilon_F)(\hbar\omega+\s\m),
	} 
	with $\m=\m_\uar-\m_\dar$. Then we obtain
	\eq{
		\Sigma^R-\Sigma^A=
		-i\d(x-x')\d(x)\frac{\tilde\l s\hbar}{32\p}J_{int}^2 A^2\sum_\s(\hbar\omega+\s\m)
		\begin{pmatrix}
			\p^2-4 & (\p-2\s)^2 \\
			(\p+2\s)^2 & \p^2-4
		\end{pmatrix}
		\\
		=
		-i\d(x-x')\d(x)\frac{\tilde\l s\hbar}{4\p}J_{int}^2 A^2\left[\hbar\omega
		\begin{pmatrix}
			\frac{\p^2}{4}-1 & \frac{\p^2}{4}+1 \\
			\frac{\p^2}{4}+1 & \frac{\p^2}{4}-1
		\end{pmatrix}
		+\p\m
		\begin{pmatrix}
			0 & -1 \\
			1 & 0
		\end{pmatrix}
		\right].
	}
	Moreover, the identity
	\eq{
		G^A_{\bar{\s}}(0,\omega') G^R_\s(0,\omega'-\omega)+G^R_{\bar{\s}}(0,\omega') G^A_\s(0,\omega'-\omega)=\left[G^A_{\bar{\s}}(0,\omega')-G^R_{\bar{\s}}(0,\omega')\right]
		\left[G^R_{\s}(0,\omega'-\omega)-G^A_\s(0,\omega'-\omega)\right],
	}
	combined with
	\eq{
		1-\tanh(x)\tanh(y)=\left(\tanh(x)-\tanh(y)\right)\coth(x-y),
	}
	yields
	\eq{
		\Sigma^K=-i\d(x-x')\d(x)\frac{\tilde\l s\hbar}{32\p}J_{int}^2 A^2\sum_\s(\hbar\omega+\s\m)
		\begin{pmatrix}
			\p^2-4 & (\p-2\s)^2 \\
			(\p+2\s)^2 & \p^2-4
		\end{pmatrix}
		\coth\left(\frac{\hbar\omega+\s\m}{2k_B T}\right).
	}
	We furthermore obtain
	\eq{
		\s_1i\Im{\Sigma^{R}}=-\frac{i\tilde\l s\hbar}{8\p}J_{int}^2A(\varepsilon_F/\hbar)^2\d(x-x')\d(x)\left[\hbar\omega
		\begin{pmatrix}
			\frac{\p^2}{4}+1 & \frac{\p^2}{4}-1 \\
			\frac{\p^2}{4}-1 & \frac{\p^2}{4}+1
		\end{pmatrix}
		+\p\m\s_3\right].
	}
	We obtain the results for the right interface by replacing $\d(x)\rightarrow\d(x-d)$ and recognizing the interface damping length
	$
	\a'=\tilde\l sJ_{int}^2A(\varepsilon_F/\hbar)^2/4\p
	$.
	Then, using a Hubbard-Stratonovich transformation we obtain the equation of motion for the classical field,
	\eq{
		\label{eq:EOM}
		\int dx'\Bigg[\hbar\omega\s_3-H_{DW}-\frac{1}{2\hbar}\s_1-\frac{1}{2\hbar}\s_1 \left(\sum_\s\Sigma^l_\s+\Sigma^b+ \Sigma^r\right)
		\Bigg]\d(x-x')\cdot
		\z^c(x',\omega)=h(x,\omega),
	}
	Between $\zeta^c$ and the conventional classical field $\zeta=(\zeta^++\zeta^-)/2$ is a factor of $\sqrt{2}$ difference, which is resolved by a redefinition of the stochastic field, resulting in a factor of $1/2$ in the fluctuation dissipation theorem \cref{eq:fdt1,eq:fdt2}.
	To obtain a Landauer-Büttiker formula we insert the stochastic equation of motion to write
	\eq{
		\int dy & \langle \6_t n_z\rangle|_{x=d,t=0}
		\approx	
		\frac{-\p}{8 s}
		\Tr\left[
		\begin{pmatrix}
			1 & 1 \\
			1 & 1
		\end{pmatrix}
		\cdot\langle\dot\z^c(d,0){(\z^{c})^\dagger}(d,0)+\z^c(d,0)\dot{(\z^{c})^\dagger}(d,0)\rangle\right]
		\\
		=&
		\frac{-1}{16 \p s \hbar}\int d\omega \int d\omega'
		\\
		&\Im\left[
		\Tr\left[
		\begin{pmatrix}
			1 & 1 \\
			1 & 1
		\end{pmatrix}
		\cdot\s_3\cdot\left\{\left(\underbrace{-i\a_G\hbar\omega+H_{DW}}_{\textrm{Transport}}-\underbrace{\frac{1}{2\hbar}\s_1\cdot\Sigma^r
		}_{\textrm{Lead damping}}
		\right)\cdot\langle\z^c(d,\omega){(\z^{c})^\dagger}(d,\omega')\rangle+\underbrace{\langle h(x,\omega){(\z^{c})^\dagger}(d,\omega')\rangle}_{\textrm{Fluctuations}}\right\}\right]\right].
	}
	We identify the total transport through the right lead as the combination of the lead damping and fluctuations.
	Neglecting the bulk to lead transport, i.e. at homogeneous temperature, we obtain the transport between the leads as
	\eq{
		I_{l\rightarrow r}=
		\frac{-1}{8 s \hbar^2}\int d\omega\Re\left\{
		\Tr\left[
		g^\dagger_{-\omega}(0)\cdot
		\begin{pmatrix}
			1 & 1 \\
			1 & 1
		\end{pmatrix}
		\cdot\s_3\cdot\left(\frac{1}{2}\s_1\Sigma^r(\omega)\cdot g_\omega(0) \cdot \s_1\cdot\sum_\s\Sigma^l_\s(\omega) \left[F^l_\s(\omega)-F^r(\omega)\right]\right)\right]\right\}.
	}
	\end{widetext}
	\bibliography{References_MagnonDW}
\end{document}